\documentclass{article}

\usepackage{arxiv}

\usepackage[utf8]{inputenc} % allow utf-8 input
\usepackage[T1]{fontenc}    % use 8-bit T1 fonts
\usepackage{hyperref}       % hyperlinks
\usepackage{url}            % simple URL typesetting
\usepackage{booktabs}       % professional-quality tables
\usepackage{amsfonts}       % blackboard math symbols
\usepackage{amssymb}
\usepackage{nicefrac}       % compact symbols for 1/2, etc.
\usepackage{microtype}      % microtypography
\usepackage{lipsum}
\usepackage{subcaption}
\usepackage{fancyhdr, lastpage}
\usepackage{wrapfig}
\usepackage{blkarray}
\usepackage{nicematrix}
\usepackage{rotating}
\usepackage{multirow}
\DeclareMathAlphabet{\pazocal}{OMS}{zplm}{m}{n}
\SetMathAlphabet\pazocal{bold}{OMS}{zplm}{bx}{n}
\newtheorem{theorem}{Theorem}

\newtheorem{definition}[theorem]{Definition}

\newtheorem{example}[theorem]{Example}

\title{Hypergraph Laplacians in Diffusion Framework}

\author{
  Mehmet Emin Aktas\\
  Department of Mathematics and Statistics\\
  University of Central Oklahoma\\
  Edmond, OK 73034 \\
  \texttt{maktas@uco.edu} \\
   \And
 Esra Akbas \\
  Department of Computer Science\\
  Oklahoma State University\\
  Stillwater, OK 74078 \\
  \texttt{eakbas@okstate.edu} \\
}

\begin{document}
\maketitle

\begin{abstract}
Networks are important structures used to model complex systems where interactions take place. In a basic network model, entities are represented as nodes, and interaction and relations among them are represented as edges. However, in a complex system, we cannot describe all relations as pairwise interactions, rather should describe as higher-order interactions. Hypergraphs are successfully used to model higher-order interactions in complex systems. In this paper, we present two new hypergraph Laplacians based on diffusion framework. Our Laplacians take the relations between higher-order interactions into consideration, hence can be used to model diffusion on hypergraphs not only between vertices but also higher-order structures. These Laplacians can be employed in different network mining problems on hypergraphs, such as social contagion models on hypergraphs, influence study on hypergraphs, and hypergraph classification, to list a few.
\end{abstract}

% keywords can be removed

\section{Introduction}

Networks are important structures used to model complex systems where interactions take place. In a basic network model, entities are represented as nodes and interaction/relations among them are represented as edges. Many different areas utilize the network data to model complex relations such as biology, chemistry, finance, and social sciences. Diffusion on networks is an important concept in network science that models how a stuff, such as information and heat, diffuses between vertices based on network topology, the pattern of who is connected to whom. For example, in a social network, modeling information diffusion can be useful in rumor controlling \cite{li2017survey}. The graph Laplacian has been used to model diffusion on basic networks that only consider pairwise relations between entities. However, as we see in different real-world applications, such as human communication, chemical reactions, and ecological systems, interactions can occur in groups of three or more nodes. They cannot be simply described as pairwise relations~\cite{battiston2020networks}, rather should be described as \textit{higher-order} interactions. For example, in ecological systems, interactions can occur in groups of three or more nodes. As another example, in a coauthorship network, nodes represent authors and edges represent coauthorship between authors. An article with three authors results in an edge between each pair of the authors. However, these edges cannot be distinguished from an edge that corresponds to an article with two authors. Hence, the rich higher-order interactions are lost in the basic network model and we need to take higher-order interactions into consideration for a more accurate representation of complex systems. As one solution to this problem, \textit{hypergraphs} are used to model complex systems \cite{zhou2006learning,huang2015scalable}. In a hypergraph, nodes again represent entities as it happens for graphs, but differently, a hypergraph has \textit{hyperedges} for higher-order interactions in the network.

Despite the extreme success of the graph Laplacian in the network mining area, there are limited studies on hypergraph Laplacians. \cite{chung1993laplacian,lu2011high,hu2015laplacian,cooper2012spectra,feng1996spectra, horak2013spectra, chan2018spectral}. Besides, existing studies mainly have three issues. Firstly, the hypergraph Laplacians in these studies are only defined for special hypergraphs, such as uniform hypergraphs, i.e., the hypergraphs with a fixed size hyperedges. Secondly, these hypergraph Laplacians neglect the relations between higher-order structures i.e., is not suitable to model diffusion on hypergraphs. Thirdly, the hypergraph Laplacians are only used for computing the graph-theoretic measures such as the average minimal cut, the isoperimetric number, the max-cut, and the independence number. These studies mostly about the spectral theory of the hypergraph Laplacians and could not find a place in the applied network science area. For example, there are no studies that employ the hypergraph Laplacian to model diffusion between higher-order structures and its applications in the network mining area. 

To address the issues discussed above, we develop new and more general hypergraph Laplacians in this paper. We first represent hypergraphs as a \textit{simplicial complex}. A simplicial complex is a topological object which is built as a union of vertices, edges, triangles, tetrahedron, and higher-dimensional polytopes, i.e. \textit{simplices}. In our representation, simplices will represent hyperedges. We then develop two hypergraph Laplacians, one is based on diffusion between fixed dimension simplices and the other is based on diffusion between all simplices. Our Laplacians do not require the hypergraph to be uniform. Our objective here is to take the relation between hyperedges, i.e. simplices, into consideration for hypergraph Laplacians, which is, for instance, crucial in modeling diffusion on hypergraphs.

The paper is formatted as follows. In Section \ref{sec:pre}, we first give the necessary preliminaries and background on graphs, hypergraphs, and Laplacians. In Section \ref{sec:hyp}, we develop two new hypergraph Laplacians to address the needs. Our final remarks with future work directions are found in Section \ref{sec:conc}.

\section{Preliminaries and Background} \label{sec:pre}
In this section, we discuss the preliminary concepts for graphs, hypergraphs, graph Laplacian and hypergraph Laplacian. We also elaborate on related work with a particular focus on the hypergraph Laplacian that uses simplicial complex.
\subsection{Graphs and Hypergraphs}

\textit{Graphs}, also called \textit{networks} in literature, are structured data representing relationships between objects \cite{aggarwal2010managing,Cook2006}. They are formed by a set of \textit{vertices} (also called nodes) and a set of \textit{edges} that are connections between pairs of vertices. In a formal definition, a network $G$ is a pair of sets $G = (V, E)$ where $V$ is the set of vertices and $E \subset V \times V $ is the set of edges of the network.

There are various types of networks that represent the differences in the relations between vertices. While in an \textit{undirected network}, edges link two vertices ${v, w}$ symmetrically, in a \textit{directed network}, edges link two vertices asymmetrically. If there is a score for the relationship between vertices that could represent the strength of interaction, we can represent this type of relationship or interactions by a \textit{weighted network}. In a weighted network, a weight function $w: E \rightarrow \mathbb{R}$ is defined to assign a weight for each edge.

A \textit{hypergraph} $H$ denoted by $H=(V,E=(e_i)_{i \in I})$ on the vertex set $V$ is a family $(e_i)_{i \in I}$ ($I$ is a finite set of indexes) of subsets of $V$ called \textit{hyperedges}. We say that a hypergraph is regular if all its vertices have the same degree and uniform if all its hyperedges have the same cardinality.

In the rest of the paper, we study weighted undirected graphs and hypergraphs unless otherwise is stated.

\subsection{Graph and Hypergraph Laplacians}

\subsubsection{Graph Laplacian}
Let $G$ be a weighted undirected graph with the vertex set $V$ and a weight function $w:E \rightarrow \mathbb{R}^{\geq 0}$. Here we assume that $w(v,v)=0$ for all $v \in V$ and if two vertices $u$ and $v$ are not adjacent, then $w(u,v)=0$. A unweighted graph can be viewed as a special weighted graph with weight 1 on all edges and 0 otherwise.

The \textit{adjacency matrix} $A$ of $G$ is defined as the $n \times n$ matrix with $A(i,j)=w(v_i,v_j)$ for $i,j \in \{1,...,n\}$ with $n$ being the number of vertices of $G$. Furthermore, let $D$ be the $n \times n$ diagonal matrix with $D(i,i)=\sum_j w(i,j)$, i.e., the weighted degree of the vertex $i \leq n$. 

The graph Laplacian, first appeared in \cite{kirchhoff1847ueber} where the author analyzed flows in electrical networks, is an operator on a real-valued function on vertices of a graph. We can define the graph \textit{Laplacian} $L$ as $L=D-A$ where $D$ is the weighted degree matrix and $A$ is the weighted adjacency matrix. It has been shown that the spectrum of Laplacian is related with many graph features such as connected components, spanning trees, centralities, and diffusion \cite{mohar1997some}. 

\subsubsection{Hypergraph Laplacians} \label{sec:hyperLap}

In this paper, we develop hypergraph Laplacians inspiring from the simplicial complex hypergraph  representation, namely the \textit{simplicial Laplacian} (or Hodge Laplacian). That is why, in this section, we explain the simplical Laplacian. For other hypergraph Laplacians, we refer readers to the following references \cite{chung1993laplacian,lu2011high,hu2015laplacian,cooper2012spectra,feng1996spectra, chan2018spectral} We start with the definition of a simplicial complex.

\begin{figure}[h!]
    \centering
     \includegraphics[width=.5\textwidth]{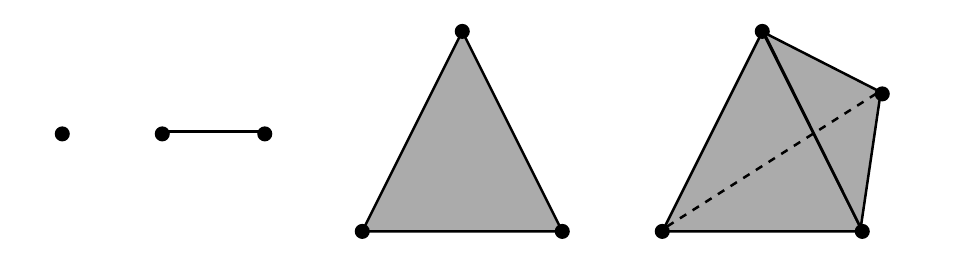}
    \caption{0-,1-,2-, and 3-simplex from left to right (borrowed from~\cite{aktas2019persistence}).}
    \label{fig:simplices}
\end{figure}

A \textit{simplicial complex} is a topological object which is built as a union of points, edges, triangles, tetrahedron, and higher-dimensional polytopes, namely \textit{simplices}.  A 0-simplex is a point, a 1-simplex is two points connected with a line segment, a 2-simplex is a filled triangle etc (see Figure \ref{fig:simplices}). 

More formally, a simplicial complex $K$ is a finite collection of simplices, i.e., points, edges, triangles, tetrahedron, and higher-dimensional polytopes, such that every face of a simplex of $K$ belongs to $K$ and the intersection of any two simplices of $K$ is a common face of both of them. In graphs, 0-simplices correspond to vertices, 1-simplices to edges, 2-simplices to triangles, and so on. We denote an $i$-simplex as $\sigma=[v_0,\dots,v_{i}]$ where $v_j \in V$ for all $j\in \{0, \dots, i\}$.

Let $S_p(K)$ be the set of all $p$-simplices of $K$. An $i$-\textit{chain} of a simplicial complex $K$ over the field $\mathbb{R}$ is a formal sum of its $i$-simplices and $i$-th \textit{chain group} of $K$ with real number coefficients, $C_i(K)=C_i(K,\mathbb{R})$, is a vector space over $\mathbb{R}$ with basis $S_i(K)$. The $i$-th \textit{cochain group} $C^i(K)=C^i(K,\mathbb{R})$ is the dual of the chain group which can be defined by $C^i(K):=\text{Hom}(C_i(K),\mathbb{R})$. Here Hom$(C_i,\mathbb{R})$ is the set of all homomorphisms of $C_i$ into $\mathbb{R}$. For an $(i+1)$-simplex $\sigma=[v_0,\dots,v_{i+1}]$, its \textit{coboundary operator}, $\delta_i:C^{i}(K) \rightarrow C^{i-1}(K)$, is defined as
$$
(\delta_i f)(\sigma)=\sum_{j=1}^{i+1}(-1)^j f([v_0,\dots,\hat{v}_j,\dots,v_{i+1}]),
$$
where $\hat{v}_j$ denotes that the vertex $v_j$ has been omitted. The \textit{boundary operators}, $\delta_i^*$, are the adjoints of the coboundary operators,
$$
\cdots C^{i-1}(K) \overset{\delta_{i+1}}{\underset{\delta_{i+1}^*}\rightleftarrows} C^i(K) \overset{\delta_{i}}{\underset{\delta_{i}^*}\rightleftarrows} C^{i+1}(K) \cdots
$$
satisfying $(\delta_i a,b)_{C^{i+1}} = (a,\delta_i^*b)_{C^i}$ for every $a \in C^{i}(K)$ and $b \in C^{i+1}(K)$, where $(\cdot ,\cdot)_{C^i}$ denote the scalar product on the cochain group. 

In \cite{horak2013spectra}, the three \textit{simplicial Laplacian operators} for higher-dimensional simplices, using the boundary and coboundary operators between chain groups, are defined as
$$\pazocal{L}_p^{\text{down}}=\delta_{p-1}\delta_{p-1}^* \text{ ~~~~~~~~~~ down Laplacian~~}$$
$$\pazocal{L}_p^{\text{up}}(K)=\delta_p^*\delta_p              \text{ ~~~~~~~~~~~~~        up Laplacian~~~~~}$$
$$~ \pazocal{L}_p(K)=\pazocal{L}_p^{\text{up}}+\pazocal{L}_p^{\text{down}} \text{   ~~~~~  Laplacian~~~~~~~~~~}$$
These operators are self-adjoint, non-negative, compact and have different spectral properties \cite{horak2013spectra}.

To make the expression of Laplacian explicit, they identify each coboundary operator $\delta_p$ with an incidence matrix $D_p$ in \cite{horak2013spectra}. The \textit{incidence matrix} $D_p \in \mathbb{R}_2^{n_{p+1}} \times \mathbb{R}_2^{n_p}$ encodes which $p$-simplices are incident to which $(p+1)$-simplices where $n_p$ is number of $p$-simplices. It is defined as
$$
D_p(i,j)= \left\{\begin{tabular}{ll}
 $1$ & if $\sigma_j^p$ is on the boundary of $\sigma_i^{p+1}$ \\
$0$  & otherwise
\end{tabular}\right.
$$
Here, we assume the simplices are not oriented. One can incorporate the orientations by simply adding ``$D_p(i,j) = -1$ if $\sigma_j^p$ is not coherent with the induced orientation of $\sigma_i^{p+1}$" in the definition if needed.

Furthermore, we assume that the simplices are weighted, i.e. there is a weight function $z$ defined on the set of all simplices of $K$ whose range is $\mathbb{R}^{+}$. Let $W_p$ be an $n_p \times n_p$ diagonal matrix with $W_p(j,j)=z(\sigma_j^p)$ for all $j \in \{1,\dots, n_p\}$. Then, the $i$-dimensional up Laplacian can be expressed as the matrix
$$\pazocal{L}_i^{\text{up}}=W_i^{-1}D_i^{T}W_{i+1}D_i.$$
Similarly, the $i$-dimensional down Laplacian can be expressed as the matrix 
$$ \pazocal{L}_i^{\text{down}}=D_{i-1}W_{i-1}^{-1}D_{i-1}^{T}W_i.$$
$\pazocal{L}_i^{\text{down}}$ is only defined for $i\geq 1$ and is equal to 0 for $i=0$. Then, to express the $i$-dimensional Laplacian $\pazocal{L}_i$, we can add these two matrices. 

\section{Hypergraph Laplacians} \label{sec:hyp}

In this paper, we develop two new hypergraph Laplacians motivating from diffusion framework. Using the simplicial Laplacian defined in Section \ref{sec:hyperLap} in diffusion has two issues. First, for a fixed simplex dimension $k$, the up simplicial Laplacian models the diffusion through only $(k+1)$-simplices and the down simplicial Laplacian only $(k-1)$-simplices. However, in the diffusion framework, a stuff on a simplex, such as heat or information, can diffuse through other simplices regardless of the dimension. For instance, in a coauthorship network, the simplicial Laplacians assume, for example, an article with three authors can affect other articles with three authors through only the articles with two or four authors. But this is not realistic since that article may also affect other articles through an article with one author as well. Second, when we use the simplicial Laplacians in modeling diffusion, we need to assume that a stuff only diffuses between $k$-simplices. However, a simplex can affect other simplices regardless of the dimension. For instance, in a coauthorship network, the simplicial Laplacians assume, for example, an article with three authors can affect only the articles with three authors. But this is again not realistic for a similar reason. 

To address these two issues, we define two new hypergraph Laplacians over simplices. The first hypergraph Laplacian allows defining diffusion between fixed dimension simplices \textit{through} any simplices, which addresses the first issue. The second hypergraph Laplacian we propose allows defining diffusion between \textit{any} simplices through any simplices, which addresses the second issue. Here, we construct these two Laplacians. 

\subsection{A hypergraph Laplacian between fixed dimension simplices} \label{sec:lapfixed}
In this section, we prose a Laplacian between a fixed dimension simplices through any simplices. Let $H$ be a hypergraph with the maximum simplex dimension $n$. In the simplicial Laplacian in Section \ref{sec:hyperLap}, the incidence matrix is only defined between $p$-simplices and $(p+1)$-simplices for $0 \leq p < n$. In order to define the Laplacian between $p$-simplices through other simplices, not only $(p-1)$- and $(p+1)$-simplices, we define a new incidence matrix as follows.

\begin{definition} \label{def:inc}
The \textit{incidence matrix} between $p$- and $r$-simplices $D_{p,r} \in \mathbb{R}_2^{n_r} \times \mathbb{R}_2^{n_p}$ for $p<r$ encodes which $p$-simplices are incident to which $r$-simplices where $n_p$ is number of $p$-simplices. It is defined as
$$
D_{p,r}(i,j)= \left\{\begin{tabular}{ll}
 $1$ & if $\sigma_j^p$ is on the boundary of $\sigma_i^r$ \\
$0$  & otherwise
\end{tabular}\right.
$$
\end{definition}

The incidence matrix in the definition above allows us to define the Laplacian between $k$-simplices through any simplices as follows.

\begin{definition}\label{def:lapkk}
Laplacian between $k$-simplices through $l$-simplices with $k\neq l$ in a hypergraph is defined as 
$$
\pazocal{L}_{k,l}=\left\{\begin{tabular}{ll}
 $W_l^{-1}D_{k,l}^{T}W_{k}D_{k,l}$ & if $k<l$ \\
$D_{l,k}W_{k}^{-1}D_{l,k}^{T}W_l$  & if $k>l$ 
\end{tabular}\right.
$$
\end{definition}
In the definition above, we follow the idea of the up and down Laplacians defined in Section \ref{sec:hyperLap}. Now, to define the hypergraph Laplacian between $k$-simplices through all simplices, we add up all the Laplacians as follows.
\begin{definition} \label{def:lapk}
Let $H$ be a hypergraph with the maximum simplex dimension $n$. Then, Laplacian between $k$-simplices through other simplices in $H$ is defined as 
$$
\pazocal{L}_{k}=\pazocal{L}_{k,0}+\pazocal{L}_{k,1}+\cdots+\pazocal{L}_{k,k-1}+\pazocal{L}_{k,k+1} + \cdots+\pazocal{L}_{k,n}
$$
for $k\in \{0,\dots,n\}$.
\end{definition}

Here we provide an example to the hypergraph Laplacian in Definition \ref{def:lapk} on a toy graph. 

\begin{example}
The simplicial complex in Figure \ref{fig:simpcomp} has four vertices (0-simplices), five edges (1-simplices) and two triangles (2-simplices).

\begin{figure}[h!]
    \centering
     \includegraphics[width=.25\textwidth]{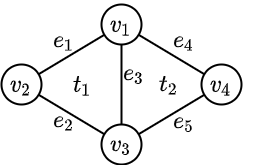}
    \caption{An unweighted simplicial complex with four vertices (0-simplices), five edges (1-simplices) and two triangles (2-simplices).}
    \label{fig:simpcomp}
\end{figure}

The corresponding incidence matrices as in Definition \ref{def:inc} are as follows.
\[
D_{0,1}=\begin{blockarray}{ccccc}
& v_1 & v_2 & v_3 & v_4 \\
\begin{block}{c(cccc)}
  e_1 & 1 & 1 & 0 & 0 \\
  e_2 & 0 & 1 & 1 & 0 \\
  e_3 & 1 & 0 & 1 & 0  \\
  e_4 & 0 & 0 & 1 & 1  \\
  e_5 & 1 & 0 & 0 & 1 \\
\end{block}
\end{blockarray}, 
D_{0,2}=\begin{blockarray}{ccccc}
& v_1 & v_2 & v_3 & v_4 \\
\begin{block}{c(cccc)}
  t_1 & 1 & 1 & 1 & 0 \\
  t_2 & 1 & 0 & 1 & 1 \\
\end{block}
\end{blockarray}, 
D_{1,2}=\begin{blockarray}{cccccc}
& e_1 & e_2 & e_3 & e_4 & e_5 \\
\begin{block}{c(ccccc)}
  t_1 & 1 & 1 & 1 & 0 & 0 \\
  t_2 & 0 & 0 & 1 & 1 & 1 \\
\end{block}
\end{blockarray}.
 \]
Then, following Definitions \ref{def:lapkk} and \ref{def:lapk}, we get the Laplacian between 0, 1 and 2-simplices as follows.
\[
\pazocal{L}_{0}=\pazocal{L}_{0,1}+\pazocal{L}_{0,2}=\begin{pmatrix} 3 & 1 & 1 & 1 \\ 1 & 2 & 1 & 0 \\ 1 & 1 & 3 & 1 \\ 1 & 0 & 1 & 2\end{pmatrix} + \begin{pmatrix} 2 & 1 & 2 & 1 \\ 1 & 1 & 1 & 0 \\ 2 & 1 & 2 & 1 \\ 1 & 0 & 1 & 1\end{pmatrix} = \begin{pmatrix} 5 & 2 & 3 & 2 \\ 2 & 3 & 2 & 0 \\ 3 & 2 & 5 & 2 \\ 2 & 0 & 2 & 3\end{pmatrix},
\]

\[
\pazocal{L}_{1}=\pazocal{L}_{1,0}+\pazocal{L}_{1,2}=\begin{pmatrix} 2 & 1 & 1 & 0 & 1 \\ 1 & 2 & 1 & 1 & 0 \\ 1 & 1 & 2 & 1 & 1 \\ 0 & 1 & 1 & 2 & 1 \\ 1 & 0 & 1 & 1 & 2\end{pmatrix} + \begin{pmatrix} 1 & 1 & 1 & 0 & 0 \\ 1 & 1 & 1 & 0 & 0 \\ 1 & 1 & 2 & 1 & 1 \\ 0 & 0 & 1 & 1 & 1 \\ 0 & 0 & 1 & 1 & 1\end{pmatrix} = \begin{pmatrix} 3 & 2 & 2 & 0 & 1 \\ 2 & 3 & 2 & 1 & 0 \\ 2 & 2 & 4 & 2 & 2 \\ 0 & 1 & 2 & 3 & 2 \\ 1 & 0 & 2 & 2 & 3\end{pmatrix},
\]

\[
\pazocal{L}_{2}=\pazocal{L}_{2,0}+\pazocal{L}_{2,1}=\begin{pmatrix} 3 & 2 \\ 2 & 3\end{pmatrix} + \begin{pmatrix} 3 &1 \\ 1 & 3\end{pmatrix} = \begin{pmatrix} 6 & 3 \\ 3 & 6\end{pmatrix}.
\]

\end{example}

We can interpret the hypergraph Laplacians for a fixed dimension as follows. For a fixed dimension $k$, the diagonal entries show the number of neighboring simplices for each $k$-simplex. For example, in the toy graph, the vertex $v_1$ has five neighboring simplices as $e_1, e_3, e_5, t_1, t_2$. That is why $\pazocal{L}_0(1,1)=5$. Furthermore, the off-diagonal entries show the number of shared neighboring simplices between $k$-simplices. For example, $v_1$ and $v_2$ share two neighboring simplices as $e_1, t_1$. That is why $\pazocal{L}_0(1,2)=\pazocal{L}_0(2,1)=2$.  

\subsection{A generalized hypergraph Laplacian} \label{sec:lapgen}
The hypergraph Laplacian in Definition \ref{def:lapk} extends the simplicial Laplacian in a way to allow the diffusion through any simplices between fixed dimensional simplices. However, this Laplacian is not able to capture the diffusion between different dimensional simplices. In order to define the generalized hypergraph Laplacian, we define a new incidence matrix that allows to encode the relation between $p$- and $r$-simplices through $q$ simplices with $p<r$ and $q \notin \{p,r\}$ as follows.

\begin{definition}\label{def:incprq}
Incidence matrix between $p$- and $r$-simplices through $q$ simplices $D_{p,r}^q \in \mathbb{R}_2^{n_r} \times \mathbb{R}_2^{n_p}$ with $p<r$ encodes which $p$-simplices are incident to which $r$-simplices through $q$-simplices where $n_p$ is number of $p$-simplices. For $q \notin \{p,r\}$, it is defined as
$$
D_{p,r}^q(i,j)= s
$$
where $s$ is the number of the $q$-simplices that are adjacent to both $\sigma_j^p$ and $\sigma_i^r$. For $q \in \{p,r\}$, we take $D_{p,r}^p=D_{p,r}^r=D_{p,r}$ as in Definition \ref{def:inc}.
\end{definition}

Now, using the incidence matrix defined above, we define a new incidence matrix between $p$- and $r$-simplices through all simplices as follows.

\begin{definition}\label{def:incpq}
Incidence matrix between $p$- and $r$-simplices through all simplices $\pazocal{D}_{p,r} \in \mathbb{R}_2^{n_r} \times \mathbb{R}_2^{n_p}$ with $p<r$ encodes which $p$-simplices are incident to which $r$-simplices through any simplex where $n_p$ is number of $p$-simplices. It is defined as
$$
\pazocal{D}_{p,r}=\sum_{i=0}^n D_{p,r}^i.
$$
\end{definition}
As the final step, we define the generalized hypergraph Laplacian between any simplices through any simplices as follows.

\begin{definition} \label{def:lapH}
Let $H$ be a hypergraph with the maximum simplex dimension $n$. Then we define the hypergraph Laplacian of $H$, $\pazocal{L}_H$, as the following block matrix
$$
\pazocal{L}_H=\left(
\begin{array}{c|c|c|c|c}
\pazocal{L}_0 & \pazocal{D}_{0,1}^T & \pazocal{D}_{0,2}^T & \cdots & \pazocal{D}_{0,n}^T \\ \hline
\pazocal{D}_{0,1} & \pazocal{L}_1 & \pazocal{D}_{1,2}^T & \cdots & \pazocal{D}_{1,n}^T \\ \hline
\pazocal{D}_{0,2} & \pazocal{D}_{1,2} & \pazocal{L}_2 & \cdots & \pazocal{D}_{2,n}^T \\ \hline
\vdots & \vdots & \vdots & \ddots & \vdots \\ \hline
\pazocal{D}_{0,n} & \pazocal{D}_{1,n} & \pazocal{D}_{2,n} & \cdots & \pazocal{L}_n
\end{array}\right)
$$
where $\pazocal{D}_{p,q}$ is the incidence matrix between $p$- and $r$ simplices through all simplices of $H$ and $\pazocal{L}_k$ is the Laplacian between $k$-simplices through other simplices of $H$.

\end{definition}

Here we continue the example in the previous section but this time show how to define the generalized hypergraph Laplacian. 

\begin{example}
In the toy graph in Figure \ref{fig:simpcomp}, the corresponding incidence matrices as in Definition \ref{def:incprq} are
\[
D_{0,1}^2=\begin{blockarray}{ccccc}
& v_1 & v_2 & v_3 & v_4 \\
\begin{block}{c(cccc)}
  e_1 & 1 & 1 & 1 & 0 \\
  e_2 & 1 & 1 & 1 & 0 \\
  e_3 & 2 & 1 & 2 & 1  \\
  e_4 & 1 & 0 & 1 & 1  \\
  e_5 & 1 & 0 & 1 & 1 \\
\end{block}
\end{blockarray}, 
D_{0,2}^1=\begin{blockarray}{ccccc}
& v_1 & v_2 & v_3 & v_4 \\
\begin{block}{c(cccc)}
  t_1 & 2 & 2 & 2 & 0 \\
  t_2 & 2 & 0 & 2 & 2 \\
\end{block}
\end{blockarray}, 
D_{1,2}^0=\begin{blockarray}{cccccc}
& e_1 & e_2 & e_3 & e_4 & e_5 \\
\begin{block}{c(ccccc)}
  t_1 & 2 & 2 & 2 & 1 & 1 \\
  t_2 & 1 & 1 & 2 & 2 & 2 \\
\end{block}
\end{blockarray}.
\]
Then, the incidence matrices as in Definition \ref{def:incpq} are
\[
\pazocal{D}_{0,1}=\begin{pmatrix} 3 & 3 & 1 & 0 \\ 1 & 3 & 3 & 0 \\ 4 & 1 & 4 & 1 \\ 1 & 0 & 3 & 3 \\ 3 & 0 & 1 & 3 \end{pmatrix}, \pazocal{D}_{0,2}=\begin{pmatrix} 4 & 4 & 4 & 0 \\ 4 & 0 & 4 & 4 \end{pmatrix}, 
\pazocal{D}_{1,2}=\begin{pmatrix} 4 & 4 & 4 & 1 & 1 \\ 1 & 1 & 4 & 4 & 4\end{pmatrix}.
\]

Finally, if we combine these incidence matrices with the hypergraph Laplacians as in Definition \ref{def:lapH}, we get the generalized Laplacian for the hypergraph in Figure \ref{fig:simpcomp} as follows
\[
\centering
\pazocal{L}_H=\left(
\begin{array}{cccc|ccccc|cc}
5 & 2 & 3 & 2 & 3 & 1 & 4 & 1 & 3 & 4 & 4 \\
2 & 3 & 2 & 0 & 3 & 3 & 1 & 0 & 0 & 4 & 0 \\
3 & 2 & 5 & 2 & 1 & 3 & 4 & 3 & 1 & 4 & 4 \\
2 & 0 & 2 & 3 & 0 & 0 & 1 & 3 & 3 & 0 & 4 \\ \hline
3 & 3 & 1 & 0 & 3 & 2 & 2 & 0 & 1 & 4 & 1 \\
1 & 3 & 3 & 0 & 2 & 3 & 2 & 1 & 0 & 4 & 1 \\
4 & 1 & 4 & 1 & 2 & 2 & 4 & 2 & 2 & 4 & 4 \\
1 & 0 & 3 & 3 & 0 & 1 & 2 & 3 & 2 & 1 & 4 \\
3 & 0 & 1 & 3 & 1 & 0 & 2 & 2 & 3 & 1 & 4 \\ \hline
4 & 4 & 4 & 0 & 4 & 4 & 4 & 1 & 1 & 6 & 3 \\
4 & 0 & 4 & 4 & 1 & 1 & 4 & 4 & 4 & 3 & 6 \\
\end{array}\right).
\]
\end{example}
As it happens in the previous hypergraph Laplacian, the diagonal entries show the number of neighboring simplices for each $k$-simplex and the off-diagonal entries show the number of the shared neighboring simplices with other simplices. The diffusion between simplices happens based on the number of the shared neighboring simplices with other simplices in the generalized Laplacian.

\section{Conclusion} \label{sec:conc}
In this paper, we develop two new hypergraph Laplacians based on diffusion framework. These Laplacians can be employed in different network mining problems on hypergraphs, such as social contagion models on hypergraphs, influence study on hypergraphs, and hypergraph classification, to list a few.

\bibliographystyle{unsrt}  
\bibliography{references}  %%% Remove comment to use the external .bib file (using bibtex).

\begin{thebibliography}{10}

\bibitem{li2017survey}
Mei Li, Xiang Wang, Kai Gao, and Shanshan Zhang.
\newblock A survey on information diffusion in online social networks: Models
  and methods.
\newblock {\em Information}, 8(4):118, 2017.

\bibitem{battiston2020networks}
Federico Battiston, Giulia Cencetti, Iacopo Iacopini, Vito Latora, Maxime
  Lucas, Alice Patania, Jean-Gabriel Young, and Giovanni Petri.
\newblock Networks beyond pairwise interactions: structure and dynamics.
\newblock {\em arXiv preprint arXiv:2006.01764}, 2020.

\bibitem{zhou2006learning}
Dengyong Zhou, Jiayuan Huang, and Bernhard Sch{\"o}lkopf.
\newblock Learning with hypergraphs: Clustering, classification, and embedding.
\newblock {\em Advances in neural information processing systems},
  19:1601--1608, 2006.

\bibitem{huang2015scalable}
Jin Huang, Rui Zhang, and Jeffrey~Xu Yu.
\newblock Scalable hypergraph learning and processing.
\newblock In {\em 2015 IEEE International Conference on Data Mining}, pages
  775--780. IEEE, 2015.

\bibitem{chung1993laplacian}
Fan Chung.
\newblock The laplacian of a hypergraph.
\newblock {\em Expanding graphs (DIMACS series)}, pages 21--36, 1993.

\bibitem{lu2011high}
Linyuan Lu and Xing Peng.
\newblock High-ordered random walks and generalized laplacians on hypergraphs.
\newblock In {\em International Workshop on Algorithms and Models for the
  Web-Graph}, pages 14--25. Springer, 2011.

\bibitem{hu2015laplacian}
Shenglong Hu and Liqun Qi.
\newblock The laplacian of a uniform hypergraph.
\newblock {\em Journal of Combinatorial Optimization}, 29(2):331--366, 2015.

\bibitem{cooper2012spectra}
Joshua Cooper and Aaron Dutle.
\newblock Spectra of uniform hypergraphs.
\newblock {\em Linear Algebra and its applications}, 436(9):3268--3292, 2012.

\bibitem{feng1996spectra}
Keqin Feng et~al.
\newblock Spectra of hypergraphs and applications.
\newblock {\em Journal of number theory}, 60(1):1--22, 1996.

\bibitem{horak2013spectra}
Danijela Horak and J{\"u}rgen Jost.
\newblock Spectra of combinatorial laplace operators on simplicial complexes.
\newblock {\em Advances in Mathematics}, 244:303--336, 2013.

\bibitem{chan2018spectral}
T-H~Hubert Chan, Anand Louis, Zhihao~Gavin Tang, and Chenzi Zhang.
\newblock Spectral properties of hypergraph laplacian and approximation
  algorithms.
\newblock {\em Journal of the ACM (JACM)}, 65(3):1--48, 2018.

\bibitem{aggarwal2010managing}
Charu~C Aggarwal and Haixun Wang.
\newblock {\em Managing and mining graph data}, volume~40.
\newblock Springer, 2010.

\bibitem{Cook2006}
Diane~J. Cook and Lawrence~B. Holder.
\newblock {\em Mining Graph Data}.
\newblock John Wiley \& Sons, 2006.

\bibitem{kirchhoff1847ueber}
Gustav Kirchhoff.
\newblock Ueber die aufl{\"o}sung der gleichungen, auf welche man bei der
  untersuchung der linearen vertheilung galvanischer str{\"o}me gef{\"u}hrt
  wird.
\newblock {\em Annalen der Physik}, 148(12):497--508, 1847.

\bibitem{mohar1997some}
Bojan Mohar.
\newblock Some applications of laplace eigenvalues of graphs.
\newblock In {\em Graph symmetry}, pages 225--275. Springer, 1997.

\bibitem{aktas2019persistence}
Mehmet~E Aktas, Esra Akbas, and Ahmed El~Fatmaoui.
\newblock Persistence homology of networks: methods and applications.
\newblock {\em Applied Network Science}, 4(1):1--28, 2019.

\end{thebibliography}


\end{document}